\documentstyle[12pt]{article}

\input psfig

\def\xslide#1#2#3#4#5#6{\centerline{\psfig
{figure=#1,height=#2,bbllx=#3bp,bblly=#4bp,bburx=#5bp,bbury=#6bp,clip=}}}

\def\half{{\textstyle{1\over2}}}

\begin{document}

$\mbox{}$
\vspace{3cm}

\begin{center}
\begin{large}   
                       
{\bf TOTAL $\gamma^{\star}\gamma^{\star}$ CROSS SECTION AND THE BFKL
POMERON} \end{large} \footnote{ This work was supported in part by the
KBN Grant No. 2P03B 083 08 (A.B. and W.C.), by the KBN Grant No. 2P03B
080 12 (W.F.), by the Stiftung f\"ur Deutsch-Polnische Zusammenarbeit
project 1522/94/LN (W.F.), and by PECO grant from the EEC Programme
``Human Capital Mobility'', Network ``Physics at High Energy
Colliders'' (Contract No. ERBICIPDCT 940613). } 
\\[12mm]

A. Bialas $^{(1)}$, W. Czyz $^{(1,2)}$, W. Florkowski $^{(2)}$

$^{(1)}$ {\it M. Smoluchowski Institute of Physics, Jagellonian University, \\
ul. Reymonta 4, PL-30-059 Krak\'ow, Poland}\\[5mm]

$^{(2)}${\it H. Niewodnicza\'nski Institute of Nuclear Physics, \\
ul. Radzikowskiego 152, PL-31-342 Krak\'ow, Poland} \\[7mm]

\end{center}

{\bf Abstract:} In the framework of the dipole picture of the BFKL
pomeron we discuss a few possibilities of calculating the total
$\gamma^{\star}\gamma^{\star}$ cross section of the virtual photons.
We argue that the existing successful fits of the dipole picture
formulae to the measured structure functions $F_2(Q^2,x)$ favor one of
the possible extensions of the dipole picture of the BFKL pomeron
formulated recently by one of us. 

\vspace{3mm}
\noindent TPJU --- 6/97 

\noindent May 1997

\newpage
\renewcommand{\headheight}{-10mm}  
\noindent{\bf 1. Introduction}
\bigskip

Testing the BFKL pomeron through collisions of tagged $e^+ e^-$ pairs
with very large momentum transfers is an attractive possibility which
has already been discussed in Refs. \cite{BHS,BBL}. Clearly, the crucial
ingredient in the expression for the $e^+ e^-$ total cross section at
fixed $Q^2_A, Q^2_B$ momentum transfers of the tagged leptons is the
total cross section of two virtual photons of ``masses'' $Q^2_A$ and
$Q^2_B$. Calculation of this cross section, $\sigma_{\gamma\gamma}$,
is basic for the content of Refs. \cite{BHS,BBL}.

Here, we present a method of calculating $\sigma_{\gamma\gamma}$,
alternative to the one presented in \cite{BHS} and \cite{BBL}, which
is an implementation of the dipole picture of the BFKL pomeron
proposed in Refs. \cite{Ma,Mb,Mc,NZ}. This implementation goes
somewhat beyond the applications of the dipole picture given in
Refs. \cite{NPR,NPRW,BP1,BP2,B1,BCF} and is based on a discussion of the 
role of highly asymmetric $q -\bar q$ configurations of light quarks in 
the onium - onium scattering at high energy within the framework of 
Mueller's QCD dipole picture \cite{B2}.

We start with the forward onium - onium amplitude for a single pomeron 
exchange, $F^{(1)}$, which we calculated {\it ab initio}. We believe that 
this new expression is better than the one used in Refs. 
\cite{BP1,BP2,B1,BCF}. Its detailed derivation is given in Appendix A and 
the result is

\begin{equation}
\label{f1}
F^{(1)}=\pi\alpha^2 r_A r_B\int{d\gamma\over 2\pi i}
e^{\Delta(\gamma)Y}\left({r_A\over
r_B}\right)^{\gamma-1}h(\gamma)\,\,.
\end{equation}

\noindent Here $\alpha$ is the strong coupling constant, $N$ is the 
number of colors, $r_A$ and $r_B$ are the transverse sizes of the two
colliding onia, $\Delta(\gamma)=\alpha N\chi(\gamma)/\pi$ where

\begin{equation}
\label{chi}
\chi(\gamma)=2\psi(1)-\psi(1-\half\gamma)-\psi(\half\gamma)
\,\,\,\,\,\, \left(\psi\equiv
{d \log\Gamma\over d\gamma}\right)
\end{equation}
and
\begin{equation}
\label{h}
h(\gamma)={4\over\gamma^2(2-\gamma)^2}\,\,.
\end{equation}

\noindent The quantity

\begin{equation}
\label{Y}
Y=\log\biggl({s\over s_0}\biggr)\,\,\, 
\end{equation}
is the total length of the dipole cascade, {\it i.e.}, the sum of the
cascade lengths of the two colliding onia.  $s$ is the total
c.m. energy of the collision and $s_0$ is the relevant scale of the
problem. $s_0$ cannot be calculated within the leading logarithmic
approximation and therefore it remains an unknown element in this
approach.  Its determination must rely on one's physical intuition and
on results of a phenomenological analysis of data.  In the present
paper we explore the consequences of the choice suggested by the
dipole picture \cite{B2}.

While (\ref{f1}) reproduces the saddle point approximation of
$F^{(1)}$ derived in \cite{Ma,Mb,Mc} (and employed in
\cite{BP1,BP2,B1,BCF}), it also contains contributions which are
neglected in the contour integral representations of $F^{(1)}$
employed in \cite{BP1,BP2,B1,BCF}. In other words, in Refs. 
\cite{BP1,BP2,B1,BCF}, these components of the integrand in the  
contour integral representation of $F^{(1)}$ which become unity at the
saddle point, are kept equal one throughout the whole contour
integration. This approximation has been corrected in our present
expression for $F^{(1)}$.

The way to employ the dipole picture to calculate the total
$\gamma^{\star}\gamma^{\star}$ cross section is, in principle,
straightforward. From $F^{(1)}$ and the well known (compare, {\it
e.g.}, \cite{NZ,BP2,BCF,BKS}) wave functions of the two photons, $A$
and $B$, of the virtual masses $Q_{A,B}$, longitudinally ($L$) or
transversely ($T$) polarized, $\Psi^{L,T}(r_{A,B},z_{A,B};Q_{A,B})$,
we obtain the forward $\gamma^{\star}\gamma^{\star}$ amplitude

\begin{equation}
\label{fgg}
F_{\gamma\gamma}=\int|\Psi^{L,T}(r_A,z_A;Q_A)|^2
|\Psi^{L,T}(r_B,z_B;Q_B)|^2
F^{(1)} d^2r_Adz_Ad^2r_Bdz_B \,\,,
\end{equation}
and the total cross section which, with our conventions, reads
\begin{equation}
\label{sgg}
\sigma_{\gamma\gamma} = 2\,{\rm Re}\, F_{\gamma\gamma}\,\,\,. 
\end{equation}

Then, in order to evaluate the integral in (\ref{fgg}), we have to
decide what to take for $Y$, the length of the cascade. According to
(\ref{Y}) this amounts to a selection of the scale $s_0$.  Two choices
were discussed in the literature.  In Ref. \cite{BHS} $s_0$ was taken
as

\begin{equation}
\label{s0a}
s_0 = c \, Q_AQ_B \,\,,
\end{equation}
with $c=100$. This apparently natural choice has an attractive feature
to be a simple analytic function of $Q_A$ and $Q_B$.  Another
possibility \cite{NPR} was to take 

\begin{equation}
\label{s0b}
s_0 = c\, Q_>^2 \,\,,
\end{equation}
where $Q_>$ is the larger of $Q_A\,,\,Q_B$. This gives 

\begin{equation}
\label{Yb}
Y = \log\biggl({1\over cx_{Bj}}\biggr) \,\,\,,
\end{equation}
a formula which provided an excellent fit to the proton structure
function at small $x_{Bj}$ \cite{NPR,NPRW}\footnote{Note that
(\ref{s0a}) and (\ref{s0b}) give rather different results for $Q^2$
dependence of $\gamma - \gamma$ cross section which should not be too
difficult to test once the relevant data are available.}.

Following \cite{B2} we observe, however, that from the point of view 
of the dipole picture neither (\ref{s0a}) nor (\ref{s0b}) is really 
satisfactory.  The point is that Eq. (\ref{f1}) refers to collisions 
of two {\it dipoles} and thus the relevant scale $s_0$ must be 
expressed in term of the parameters characterizing these dipoles 
({\it i.e.}, longitudinal momenta $z_A, \, z_B$ and transverse 
sizes $r_A\,, r_B$) rather than $Q_A$ and $Q_B$. This is clear
if one observes that $Q_A$ and $Q_B$ are not even defined in 
(\ref{f1}). The possible choices of $s_0$ consistent with the dipole 
picture were discussed in \cite{B2}, where also a definite formula 
for $Y$ was suggested. In the present paper we explore consequences 
of this choice for $\sigma_{\gamma\gamma}$ and compare it with the 
results following from (\ref{s0a}) and (\ref{s0b}).  We hope that our 
results shall be useful in testing the validity of the dipole picture 
approach in the small $x_{Bj}$ physics.

In the next Section we present our formulae for
$\sigma_{\gamma\gamma}$ following from $Y$ worked out in
Ref. \cite{B2}. In Section 3 we give our numerical results and their
discussion.  Section 4 contains the conclusions. Appendix A presents a
detailed derivation of the formula (\ref{f1}), in Appendix B we
calculate $\sigma_{\gamma\gamma}$ for three factorizable forms of $Y$
defined by: (\ref{taud}), (\ref{s0a}), and (\ref{s0b}). The case
(\ref{s0a}) gives $\sigma_{\gamma\gamma}$ of Ref. \cite{BHS} whereas
the last one, (\ref{s0b}), results in $\sigma_{\gamma\gamma}$ which
follows from the version of the dipole picture implemented in
Ref. \cite{NPR}.

\bigskip
\noindent{\bf 2. The total $\gamma^{\star} - \gamma^{\star}$ 
cross section}
\bigskip

In this Section we derive the formulae for the total cross section of
two virtual gammas using the formula (\ref{f1}) for {\it onium} - {\it
onium} forward amplitude derived in Appendix A, with $Y$ taken from
Ref. \cite{B2}

\begin{equation}
\label{Ytau}
Y = y_A + y_B = 
\log\left({sz^<_Az^<_B r^2_Ar^2_B\over c\tau^2_{int}}\right)\,,
\end{equation}
where $c$ is the arbitrary constant of the leading log approximation,
$s=4E_AE_B$ is the square of the total c.m. energy of the colliding
virtual photons,

\begin{equation}
\label{z}
z^< = 
\left\{
\begin{array}{c}
z  \hspace{1.25cm} \mbox{if $z\leq \half$} \\
1-z \hspace{0.5cm} \mbox{if $ z \geq \half$}, 
\end{array}
\right.
\end{equation}
and
\begin{equation}
\label{tauc}
\tau_{\rm int}={\rm const}\, r_{>}
\end{equation}
where $r_{>}$ is the larger of $r_A$ and $r_B$.  $\tau_{\rm int}$ is
interpreted (see \cite{B2}) as the time needed for the exchanged 
gluons to travel the necessary distance in the transverse space.

We will confront the results obtained with (\ref{tauc}) with the ones 
one gets replacing (\ref{tauc}) by a symmetric expression

\begin{equation}
\label{taud}
\tau_{\rm int}^2 \to \tau^2= r_A r_B
\end{equation}
which leads to formulae close to the ones advocated in Refs. 
\cite{BHS,BBL}.

To get $\gamma^{\star} - \gamma^{\star}$ amplitude we employ now the 
wave functions of the virtual photons, $\Psi(r_A,z_A;Q_A), 
\Psi(r_B,z_B;Q_B)$, and calculate $F_{\gamma\gamma}$ of (\ref{fgg}), 
where for the transverse ($T$) and longitudinal ($L$) photons we have 
(compare \cite{NZ,BP2,BCF,BKS})

\begin{equation}
\label{phi}
|\Psi^{T,L}(r,z;Q)|^2=\Phi^{T,L}(r,z;Q)=
{N\alpha_{em}e^2_f\over\pi^2}W^{T,L}(r,z;Q)\,\,,
\end{equation}

\begin{equation}
\label{wt}
W^T(r,z;Q) = \half[z^2+(1-z)^2]{\hat Q}^2 
K^2_1({\hat Q}r)\,\,,
\end{equation}

\begin{equation}
\label{wl}
W^L(r,z;Q)=2z(1-z)\hat Q^2K^2_0(\hat Qr)\,\,, 
\end{equation}

\noindent where $\hat Q=\sqrt{z(1-z)} Q,\, \alpha_{em}=1/137$ and 
$e^2_f=2/3$ (the sum of the squares of the charges of three quarks).

Inserting (\ref{Ytau}) into (\ref{f1}) and employing (\ref{fgg})
we obtain for the total $\gamma^{\star}\gamma^{\star}$ cross section
\begin{eqnarray}
\label{sgg1}
\sigma_{\gamma\gamma} & = & 4(2\pi)^3\alpha^2\int^{\infty}_0 dr_A 
r^2_A\int^{\half}_0 d z_A \Phi^{T,L}(z_A,r_A;Q_A) \nonumber \\
&\times& \int^{\infty}_0 dr_B r^2_B\int^{\half}_0 dz_B 
\Phi^{T,L}(z_B, r_B;Q_B)
\int{d\gamma\over2\pi i}e^{\Delta(\gamma)Y}
\biggl({r_A\over r_B}\biggr)^{1-\gamma}h(\gamma)\,\,,
\end{eqnarray}
where $Y=\log{\xi}$ and  $\xi=(sz^<_Az^<_B r^2_A r^2_B/(c\tau^2_{\rm int}))$.
Note that since $\Phi^T$ and $\Phi^L$ are invariant against the
replacements: $z_{A,B}\to (1-z_{A,B})$ and $(1-z_{A,B})\to z_{A,B}$,
we can drop the $^<$ superscripts and integrate over $z$'s as follows
$\int^1_0 dz \,\,\to \,\, 2\int^{1/2}_0 dz\,\,$.

The expression (\ref{sgg1}) is comparatively easy to evaluate when, 
as in (\ref{taud}), $\tau^2_{\rm int}\to \tau^2=r_Ar_B$  (the case 
close to the one of Refs. \cite{BHS,BBL}) because the integrals 
factorize into integrals over the $A$ and $B$ variables, the
integrations over $r_A$ and $r_B$ can be done analytically and we 
obtain

\begin{equation}
\label{sgg2}
\sigma_{\gamma\gamma}={8\over\pi}(\alpha N\alpha_{em}e^2_f)^2
{1\over Q_AQ_B}
\int{d\gamma\over2\pi i}\biggl({s\over cQ_AQ_B}\biggr)^{\Delta(\gamma)}
\biggl({Q_B\over Q_A}\biggr)^{1-\gamma}h(\gamma)H^{L,T}(\gamma)\,\,,
\end{equation}
where
$H^{L,T} = 4^{\Delta(\gamma)} Z^{L,T}(\gamma) S^{L,T}(\gamma)
Z^{L,T}(2-\gamma) S^{L,T}(2-\gamma)$ is the product of the functions
of $\gamma$ defined in the Appendix B.

In the case of (\ref{tauc}), however, there is no factorization and
one has to face a 5-dimensional integration with one of the integrals
being a contour integral in the complex plane $\gamma$ (along a
straight line parallel to the imaginary axis). This forces us to use a
numerical method for the evaluation of Eq. (\ref{sgg1}). It turns out
(see the discussion of our results below) that in the very high energy
limit ($s/c$ very large) one can safely use the saddle point
approximation for the contour integral

\begin{equation}
\label{sp}
\int{d\gamma\over2\pi i}e^{\Delta(\gamma)Y}
\biggl({r_A\over r_B}\biggr)^{1-\gamma}h(\gamma)
=\half\sqrt{2a_{\xi}\over\pi}h(\gamma_0)\xi^{\Delta_p}e^{-\half
a_{\xi}\log^2(r_A/r_B)}
\,\,,
\end{equation}
with $Y=\log\xi$, $\xi = s/s_0$, with $s_0$ as the case might be
(see above), $a_{\xi} = [7\alpha N\zeta(3)\log(\xi)/\pi]^{-1}$  and
$\Delta_p={\alpha N\over \pi}\chi(1)$. $\gamma_0=1-a_{\xi}\log(r_A/r_B)$
is the saddle point which, in the limit $s/c \to \infty$, equals 1.
Then the numerical integration reduces to 4 dimensions. 
Note that (\ref{sp}) exhibits the source of the substantial difference
in dependences on $Q_B/Q_A$ following from (\ref{s0a}) and (\ref{s0b}),
see Fig. 2. This difference sits in the $\xi^{\Delta_p}$ factor:

\begin{equation}
\label{dxi}
\left( {s \over c Q_A Q_B }\right)^{\Delta_p} \,\,\,\,\,\,
\hbox{against} \,\,\,\,\,\,
\left( {s \over c Q_>^2 }\right)^{\Delta_p}.
\end{equation}

\bigskip
\noindent{\bf 3. Numerical results and  discussion}
\bigskip

We considered 4 cases of $Y = \log(s/s_0)$ and calculated the corresponding
cross sections:
\begin{itemize}
\item[(a)] The case of Eq. (\ref{s0a}), $s_0 = c_{(a)} Q_A Q_B$,
employed in Refs. \cite{BHS,BBL},
\item[(b)] the case of Eq. (\ref{s0b}), $s_0 = c_{(b)} Q_>^2$,
employed in Ref. \cite{NPR},
\item[(c)] the case of Eq. (\ref{tauc}), $s_0 = (  c_{(c)} r_>^2
/(z_A^< z_B^< r_A^2 r_B^2))$, discussed in Ref. \cite{B2}, 
\item[(d)] the case of Eq. (\ref{taud}), $s_0 = (  c_{(d)} 
/(z_A^< z_B^< r_A r_B))$, discussed also in Ref. \cite{B2}. 
\end{itemize}
Comments: As shown in Appendix B, in the Case (a) we obtain the same
formula for $\sigma_{\gamma \gamma}$ as in Ref. \cite{BHS}. Also let
us note that, when $Q_A=Q_B$ and the arbitrary constants are set to
the same value $c_{(a)} = c_{(b)} = c$, the $\sigma$'s for Case (a)
and Case (b) are identical.

The Cases (a) --- (d) were calculated in the saddle point approximation
given by the formula (\ref{sp}) and subsequent 4-dimensional
integration. In Cases (a), (b) and (d), we checked the accuracy of
this procedure calculating $\sigma_{\gamma\gamma}$ analytically up to
the final contour integration over $\gamma$ which was done with the
help of {\small MATHEMATICA}. It turned out that the results of these
two procedures agree to within 15 percent.

\bigskip
In order to exhibit the asymmetries when $Q_A\not=Q_B$ we introduced
the asymmetry parameter, $\zeta$, defined as 

\begin{equation}
\label{zeta}
\zeta = {Q_B \over Q_A}.
\end{equation}
From the asymptotic forms (\ref{b7}) and (\ref{b10}) we see that the
asymmetry in $Q_A,Q_B$ in Cases (a) and (d) is given approximately by
the factor

\begin{equation}
\label{factor}
e^{-\half
a_{\xi}\log^2(\zeta)}.
\end{equation}
In the asymmetric cases, especially in Case (b), this estimate is
not good enough.

Clearly, the choices of the values of the arbitrary constant $c$ 
involved in all $Y$'s discussed in this paper are very important
in determining the size of the cross section. They can either be fitted 
to experimental results (compare Ref. \cite{NPR}) or set following some 
prejudices of the authors (compare, {\it e.g.}, Ref. \cite{BHS}): 
in Ref. \cite{BHS}

\begin{equation}
\label{ca}
c=c_{(a)}=100, \,\,\,\xi=\xi_{(a)}={s\over c_{(a)}Q_AQ_B}\,\,
\end{equation}
and in Ref. \cite{NPR}\footnote{The authors are grateful to R. Peschanski 
for providing them with $c_{(b)}$ of (\ref{cb}) which gives the fit of 
Ref.  \cite{NPR}.}
\begin{equation}
\label{cb}
c=c_{(b)}=0.57,\,\,\,\xi=\xi_{(b)}={s\over c_{(b)}Q^2_>}\,\,.
\end{equation}
The constants $c$ for Cases (c) and (d) were set to fit the $\sigma$'s
of Cases (b) and (a), respectively, for $Q_A=Q_B=4$ GeV and $\sqrt{s}
=$ 200 GeV. They come out to be: $c_{(c)}=$ 0.0055, and $c_{(d)}=$ 2.5.

   To estimate the role of $c$'s it is enough to use the asymptotic 
formula (\ref{b10}) for $Q_A = Q_B$.  We get

\begin{equation}
\label{ratio1}
{\sigma_{\gamma\gamma}(a)\over\sigma_{\gamma\gamma}(b)} =
\left({\xi_{(a)} \over \xi_{(b)}}\right)^{\Delta_p}
\sqrt{a_{\xi_{(a)}} \over a_{\xi_{(b)}}}
\end{equation}
where $a_{\xi}$ is given below (\ref{sp}). Since, in the limit
$s\to\infty$, ${\cal {O}}(\log\xi_{(a)})={\cal {O}}(\log\xi_{(b)})$,
we have approximately

\begin{equation}
\label{ratio2}
{\sigma_{\gamma\gamma}(a)\over\sigma_{\gamma\gamma}(b)}=
\biggl({c_{(b)}\over c_{(a)}}\biggr)^{\Delta_p}.
\end{equation}

In Fig. 1 we present the $\sigma_{\gamma\gamma}$'s for Cases (a) ---
(d), for $Q_A=Q_B$, setting the strong coupling constant $\alpha$ = 0.11,
hence $\Delta_p$ = 0.3. The values of the constants $c$ were taken as 
in Ref. \cite{BHS} ($c_{(a)}$ = 100) and \cite{NPR}  ($c_{(b)}$ = 0.57).
The resulting cross sections differ appreciably, consistently with
Eq.  (\ref{ratio2}). The dipole model results, (c) and (d), were fitted
to the predictions of (b) and (a), respectively, at the point
$Q^2$ = 16 GeV$^2$ and $\sqrt{s}$ = 200 GeV. One sees that they follow
closely the results of (a) and (b) for all considered values of
$Q^2$ and  $\sqrt{s}$.

\begin{figure}[h]
\xslide{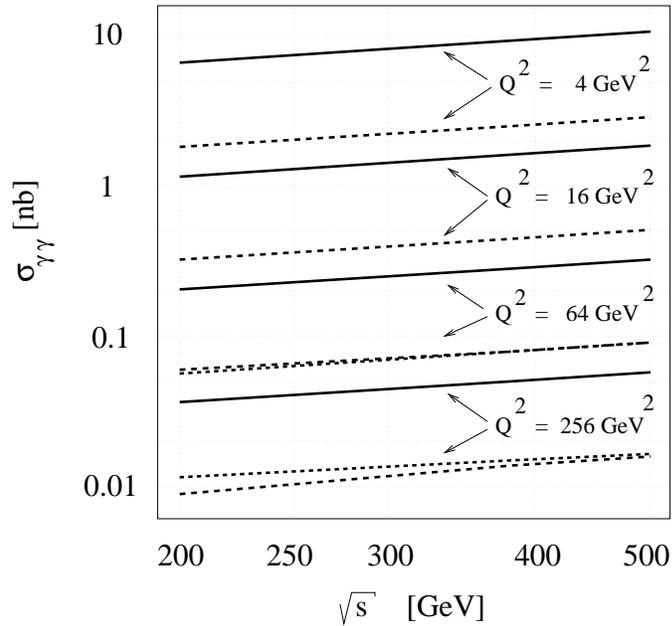}{11cm}{19}{139}{563}{742}
\caption{\small Total cross section in the dominant $TT$ channel
for $Q_A = Q_B = Q$. The solid lines represent the results for Cases
(b) and (c), whereas the dashed lines show Cases (a) and (d). Using
the values of the scale parameter $c$ given in the text, one finds
that Cases (b) and (c) coincide in the wide range of $Q^2$.  The
results for Cases (a) and (d) slightly differ for very large $Q^2$.}
\label{fig:1}
\end{figure}

\newpage
In Fig. 2 the dependence on the asymmetry parameter $\zeta = Q_B/Q_A$
is plotted. Other parameters are chosen as in Fig. 1. One sees that
the $\zeta$-dependence is almost identical for two versions of the dipole
model ((c) and (d)) and the symmetric proposal of \cite{BHS,BBL}. The
case (b) differs significantly, however, from the others, giving a
much stronger dependence on $\zeta$.

\begin{figure}[h]
\xslide{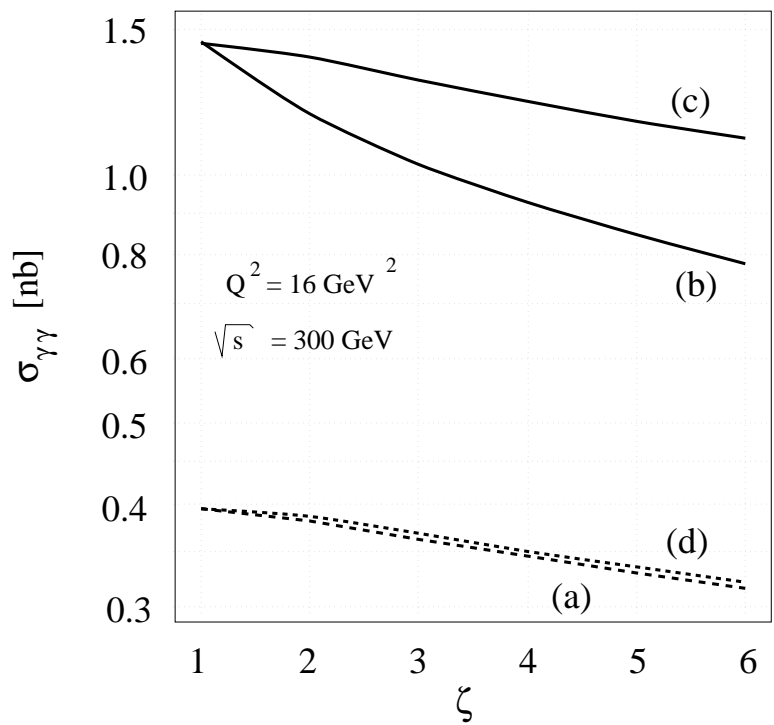}{13cm}{19}{139}{563}{742}
\caption{\small Total cross section in the dominant $TT$ channel 
for $1 \le \zeta \le 6$.} 
\label{fig:2}
\end{figure}

\newpage
\noindent{\bf 4. Conclusions}
\bigskip

The predictions of the dipole model for the photon - photon
cross section depend strongly on the scale determining  the length
of the dipole cascade in the incident photons. The scale suggested
previously \cite{BHS,BBL} gives substantially smaller cross section
than the one suggested by a fit of the dipole model results to
the proton structure function \cite{NPR}. On the other hand, the
dependence of the cross section on the ratio $Q_B/Q_A$ for the two 
colliding photons turned out to be the same for the two extreme
cases of the dipole model ((c), (d)) suggesting that it hardly
depends on the details of the model.

We conclude that future measurements of $\gamma^{*} \gamma^{*}$ cross
section may be useful in determining the length of the dipole (gluon)
cascade but, probably, not very helpful in understanding the details
of the dipole-dipole interaction. This makes rather urgent the need of
determining the relevant scales from the higher-order perturbative
calculations.

\bigskip
\noindent{\bf Acknowledgment:} The authors thank W. Broniowski for
his helpful suggestions concerning the usage of the Monte-Carlo
method.

\bigskip
\noindent{\bf Appendix A. \\ Forward onium - onium amplitude including
the light-cone momentum fractions dependencies}
\bigskip

In this Appendix we give a direct derivation of the formula for
$F^{(1)}=\int d^2b F^{(1)}(b)$ in the form of a contour integral.
This derivation accommodates the option that the dipole - dipole cross
section takes the asymptotic form only above some energy $E_0$ of the
slowest gluon.

We follow the notation of the main text: the subscripts $A$ and $B$
refer to one of the virtual photons. For example, $x_A$ is the
transverse size of a dipole originating from the virtual photon
$A$. Also $x_<$ denotes the smaller of the two sizes $x_A$ and
$x_B$. As before, $r_{A,B}$ and $z_{A,B}$ are the arguments of the
photon wave functions.  Remember, however, that $z^<_{A,B}$ are
defined by (\ref{z}). Consulting of Ref. \cite{B2} may also be helpful.

We start from the formula (3) of Mueller's paper \cite{Mc} which we 
write in the form

\begin{equation}
\label{a1}
F^{(1)}= \pi \alpha^2\int{dx_A\over x_A}{dx_B\over x_B}d^2s\sigma(x_A,x_B)
n(r_A,x_A,y_A,s)n(r_B,x_B,y_B,b-s)\,\,, 
\end{equation}
where in the limit of high energy
\begin{equation}
\label{a2}
\sigma(x_A,x_B)=4\int{dl\over l^3}[1-J_0(x_Al)][1-J_0(x_Bl)]=
x^2_<[1+\log(x_>/x_<)]\,\, ,
\end{equation}
and
\begin{equation}
\label{a3}
n(r,x,y,s)={1\over\pi^2}\int{d\gamma\over2\pi i}e^{\Delta(\gamma)y}
(1-\gamma)^2\biggl({r\over x}\biggr)^{\gamma}I(r,x,s,\gamma)\,\,,
\end{equation}
with
\begin{equation}
\label{a4}
I(r,x,s,\gamma)=\int d^2w \,
\left(|s+\half x-w||s-\half x-w|\right)^{\gamma-2}
\left(|\half r-w||-\half r-w|\right)^{-\gamma}\,\,\,.
\end{equation}

   Following Mueller we also find (see \cite{B2})
\begin{equation}
\label{a5}
y=\log\biggl({z_<\over z_0}\biggr)\,\,\,, 
\end{equation}
where $z_<$ is defined in (\ref{z}) and $z_0$ is the minimal fraction
of the onium energy which can be carried by a dipole in the wave
function (actually the minimal energy carried by one of the gluons
forming a dipole). Mueller does not explain why the integration must
stop at some $z_0$ (except for the obvious fact that for $z_0=0$ the
formula diverges). It is clear, however, that without understanding
this problem there is no chance to estimate reasonably $z_0$.
Following the arguments of \cite{B2} ({\it e.g.}, Eqs. (18) and (21)) 
we take 

\begin{equation}
\label{a6}
z_0={{\rm const}\,r_>\over r^2p^+}={p^+_0\over p^+}\,\,\,,
\end{equation}
where $p^+_0=2E_0$, $E_0$ being the energy of the slowest gluon.

When all this is substituted into (\ref{f1}) we obtain

\begin{equation}
\label{a7}
F^{(1)} = {\alpha^2\over\pi^3} 
\int {d\gamma_A\over2\pi i} {d\gamma_B\over2\pi i} 
\left({z^A_<p^+_Ar^2_A\over r_>}\right)^{\Delta(\gamma_A)}
\!\!(1-\gamma_A)^2
\biggl({z^B_<p^+_Br^2_B\over r_>}\biggr)^{\Delta(\gamma_B)}
\!\!(1-\gamma_B)^2 \Omega(r_A,r_B,\gamma_A,\gamma_B)\,\,\,,
\end{equation}
where
\begin{equation}
\label{a8}
\Omega(r_A,r_B,\gamma_A,\gamma_B)= \int{dx_A\over x_A}{dx_B\over x_B}
\sigma(x_A,x_B)\biggl({r_A\over x_A}\biggr)^{\gamma_A}
\biggl({r_B\over x_B} \biggr)^{\gamma_B}
\tilde I(x_A,x_B,r_A,r_B,\gamma_A,\gamma_B)\,\,,
\end{equation}
with
\begin{eqnarray}
\label{a9}
& &\!\!\!\!\!\!\!\!\!\!\!\!\!\!\!
\tilde I(x_A,x_B,r_a,r_B,\gamma_A,\gamma_B)=\int d^2b \, d^2s \,
d^2w \, d^2u \nonumber \\
&\times&\,\,\left(|b-s+\half x_B -u||b-s-\half x_B-u|\right)^{\gamma_B-2}
\left(|\half r_B-u||-\half r_B-u|\right)^{-\gamma_B} \nonumber \\
&\times&\,\,\left(|s+\half x_A-w||s-\half x_A-w|\right)^{\gamma_A-2}
\left(|\half r_A-w||-\half r_A-w|\right)^{-\gamma_A}\,\,\,.
\end{eqnarray}
After the change of variables, $s-w\to s,\,\,b-s-u\to b$, (\ref{a9})
factorizes:

\begin{equation}
\label{a10}
\!\!\!\!\!\!\!\!\!\!\!\!\!\!\!
\tilde I(x_A,x_B,r_A,r_B,\gamma_A,\gamma_B)=I(x_B,2-\gamma_B)
I(x_A,2-\gamma_A)I(r_A,\gamma_A)I(r_B,\gamma_B)\,\,,
\end{equation}
where
\begin{equation}
\label{a11}
I(x,\lambda)=\int d^2s(|s+\half x||s-\half x|)^{-\lambda}\,\, .
\end{equation}

$I(x,\lambda)$ can be calculated using the technique of Mueller. Here
we give only the result.  It reads 

\begin{equation}
\label{a12}
I(x,\lambda)=\pi
x^{2(1-\lambda)} H(\lambda)\,\,,
\end{equation}
where
\begin{equation}
\label{a13}
H(\lambda)={\Gamma^2(1-\half\lambda)\Gamma(\lambda-1)\over
\Gamma^2(\half\lambda)\Gamma(2-\lambda)}\,\,. 
\end{equation}
Consequently, we can write

\begin{equation}
\label{a14}
\Omega(r_A,r_B,\gamma_A,\gamma_B)
=\pi^4 H(2-\gamma_A)H(2-\gamma_B)H(\gamma_A)H(\gamma_B)
r^{2-\gamma_A}_A r^{2-\gamma_B}_B\omega(\gamma_A,\gamma_B)\,\,,
\end{equation}
where
\begin{equation}
\label{a15}
\omega(\gamma_A,\gamma_B)=\int^{\infty}_0{dx_A\over x_A}{dx_B\over x_B}
x^{\gamma_A-2}_A x^{\gamma_B-2}_B\sigma(x_A,x_B) \,\, .
\end{equation}

Integration over $x_A$ and $x_B$ is tedious but we do it explicitly
\begin{eqnarray}
\label{a16}
\omega(\gamma_A,\gamma_B) &=& \int^{\infty}_0{dx_A\over x_A}
x^{\gamma_A-2}_A 
\left[\int^{x_A}_0{dx_B\over x_B}(x_B)^{\gamma_B}(1+\log{x_A\over
x_B}) \right. \nonumber \\
&+& \left. x^2_A\int^{\infty}_{x_A}{dx_B\over x_B}(x_B)^{\gamma_B-2}
(1+\log{x_B\over x_A})\right]\,\,.
\end{eqnarray}

\noindent The integrals over $dx_B$ are well defined and are:

\begin{equation}
\label{a17}
\int^{x_A}_0{dx_B\over x_B}(x_B)^{\gamma_B}
\left(1+\log{x_A\over x_B}\right)=
x^{\gamma_B}_A{1+\gamma_B\over(\gamma_B)^2}\,\,,
\end{equation}
\begin{equation}
\label{a18}
x^2_A\int^{\infty}_{x_A}{dx_B\over x_B}(x_B)^{\gamma_B-2}
\left(1+\log{x_B\over x_A}\right)
=x^{\gamma_B}_A{3-\gamma_B\over(2-\gamma_B)^2}\,\,.
\end{equation}
When this is substituted into (\ref{a16}) and integrated over $dx_A$ 
we obtain 

\begin{equation}
\label{a19}
\omega(\gamma_A,\gamma_B)=h(\gamma_B)\int^{\infty}_0{dx_A\over x_A}
x_A^{\gamma_A+\gamma_B-2}=
{h(\gamma_B)\epsilon^{\gamma_A+\gamma_B-2}\over
2-\gamma_A-\gamma_B}\,\,,
\end{equation}
where we have regularized the integral by taking the lower limit to be
$\epsilon$.  $h(\gamma_B)$ is given by 

\begin{equation}
\label{a20}
h(\gamma_B)={4\over(2-\gamma_B)^2(\gamma_B)^2}\,\, .
\end{equation}
Inserting (\ref{a13}) and (\ref{a19}) into (\ref{a14}) we obtain 

\begin{equation}
\label{a21}
(1-\gamma_A)^2(1-\gamma_B)^2\Omega (r_A,r_B,\gamma_A,\gamma_B)
=\pi^4r_A^{2-\gamma_A}r_B^{2-\gamma_B}{h(\gamma_B)\epsilon^{\gamma_A
\gamma_B-2}\over 2-\gamma_A-\gamma_B}\,\, .
\end{equation}
Now, it turns out that the result for $F^{(1)}$ depends on the order
of integration over $\gamma$'s. The integral over $\gamma _B$ in
(\ref{a7}) can be calculated just by taking the residue at the pole
$(2-\gamma_A-\gamma_B)^{-1}$. We obtain 

\begin{equation}
\label{a22}
F^{(1)}=\pi\alpha^2r_Ar_B\int{d\gamma\over2\pi i}e^{\Delta(\gamma)Y}
\biggl({r_A\over r_B}\biggr)^{1-\gamma}h(\gamma)\,\,\,,
\end{equation}
where

\begin{equation}
\label{a23}
Y=\log(p_A^+p_B^+z_<^Az_<^Br^2_</c)=
\log(sz_<^Az_<^Br^2_</c)\,\,,
\end{equation}
and $s=p_A^+p_B^+$ is the total c.m. energy squared. The integration
over $\gamma_A$ can be similarly executed and one obtains for
$F^{(1)}$ again the result (\ref{a22}) but with $r_A,r_B\to r_B,r_A$
exchanged.  However, (\ref{a22}) is in fact invariant under this 
exchange (see, {\it e.g.}, the asymptotic form of $F^{(1)}$ 
(\ref{sp}) which exhibits such an invariance).

\bigskip
\noindent{\bf Appendix B}
\bigskip

As it was pointed out in \cite{B2}, the dipole picture of BFKL 
\cite{Ma,Mb,Mc} gives some freedom of choice of the specific 
realizations of $Y$ in (\ref{f1}). In this Appendix we give
derivations of $\sigma_{\gamma\gamma}$ for two cases of the explicitly
symmetric (with respect to the virtual photons A and B) expressions
for $Y$.

  We start with the case of $\tau^2_{\rm int}$ given by Eq. (\ref{taud}).
Now $ F^{(1)}$ becomes

\begin{equation}
\label{b1}
F^{(1)}=\pi{\alpha}^2 r_Ar_B\int{d\gamma\over2\pi i}(s r_Ar_B z^<_A
z^<_B/c)^{\Delta(\gamma)}
\biggl({r_A\over r_B}\biggr)^{\gamma-1}h(\gamma)\,\,\,,
\end{equation}
From (\ref{fgg}), (\ref{phi}), (\ref{wt}) and (\ref{wl}) it follows 
that the integrations over $r_A$, $r_B$, $z_A$, and $z_B$ factorize.  
The integrals over $r_A$ and $r_B$ can be done with the help of the 
formula 6.576/4 of Gradstein and Ryzhik \cite{GR},
and one obtains after a few straightforward operations

\begin{eqnarray}
\label{b2}
\sigma_{\gamma\gamma} &=& {8\over\pi}(\alpha N\alpha_{em}e^2_f)^2
{1\over Q_AQ_B}
\int{d\gamma\over2\pi i}h(\gamma)\biggl({4 s\over
c Q_A Q_B}\biggr)^{\Delta(\gamma)}
\biggl({Q_B\over Q_A}\biggr)^{1-\gamma} \nonumber \\
&\times& Z^{L,T}_A(\gamma) S^{L,T}_A(\gamma) Z^{L,T}_B(2-\gamma)
S^{L,T}_B(2-\gamma)\,\,\,,
\end{eqnarray}
where

\begin{equation}
\label{b3}
S^T(\gamma)={4-\gamma_{-}\over 2-\gamma_{-}}S^L(\gamma)\,\,\, , \,\,\,
S^L(\gamma)={\Gamma^4(2-\half\gamma_{-})\over\Gamma(4-\gamma_{-})}
\,\,,
\end{equation}

\begin{equation}
\label{b4}
Z^T(\gamma)=\int^{\half}_0
dz[z^2+(1-z)^2]z^{\half\gamma_{+}-1}(1-z)^{\half\gamma_{-}-1}\,\,,
\end{equation}

\begin{equation}
\label{b5}
Z^L(\gamma)=4\int^{\half}_0 dz 
z^{\half\gamma_{+}}(1-z)^{\half\gamma_{-}}
\,\,,
\end{equation}
and

\begin{equation}
\label{b6}
\gamma_{\pm}=\gamma\pm\Delta(\gamma) \,\,. 
\end{equation}

It is also of interest to have the asymptotic ({\it i.e.}, 
$s/c\to\infty$) expression for $\sigma_{\gamma\gamma}$.  This can be 
done through the saddle point approximation around $\gamma=1$ 
(from (\ref{chi}) we have that $\chi'(1)=0$). We obtain 

\begin{equation}
\label{b7}
\sigma_{\gamma\gamma}={16\over\pi}(\alpha
N\alpha_{em}e^2_f)^2{1\over Q_AQ_B}
\sqrt{2a_{\xi}\over\pi}\xi^{\Delta_p}e^{-\half a_{\xi}\log^2(Q_B/Q_A)}
[Z^{L,T}(1)S^{L,T}(1)]^2\,\,,
\end{equation}
where

\begin{equation}
\xi={4 s\over cQ_AQ_B},\,\,\,\,
\Delta_P=\Delta(\gamma=1),\,\,\,\,
a_{\xi}=[7\alpha N\zeta(3)\log(\xi)/\pi]^{-1}\,, \nonumber
\end{equation}

\begin{equation}
S^T(1) = S^L(1) {3+\Delta_P\over 1+\Delta_P}\,\,\,\,,\,\,\,\,
S^L(1) = {\Gamma^4({3\over2}+\half\Delta_P)
\over\Gamma(3+\Delta_P)}\,\,, \nonumber
\end{equation}

\begin{equation}
 Z^T(1)=\int^{\half}_0 dz[z^2+(1-z)^2]z^{\half(\Delta_P-1)}
(1-z)^{-\half(\Delta_P+1)}\,\,, \nonumber
\end{equation}

\begin{equation}
Z^L(1)=4\int^{\half}_0 dz 
z^{\half(1+\Delta_P)}(1-z)^{\half(1-\Delta_P)}\,\,.
\end{equation}

To have a direct comparison of the dipole picture formulae with those
of Refs. \cite{BHS,BBL} and with Ref. \cite{NPR}, we employ the forms 
of $Y$ which do not depend on $r$'s and $z$'s ((\ref{Y}) and (\ref{s0a}),
(\ref{Y}) and (\ref{s0b})). Now all integrals over $r_A$ , $r_B$, 
$z_A$, $z_B$ can be done analytically employing the formula 6.576/4 
of Ref. \cite{GR} for the integrals over $r$'s, and the integrals over
$z$'s are now simple Euler $\beta$ functions. We obtain 

\begin{equation}
\label{b8}
\sigma_{\gamma\gamma} = {32\alpha^2 N^2\alpha^2_{em}(e^2_f)^2 \over 
\pi Q_AQ_B}
\int{d\gamma\over2\pi i}\xi^{\Delta(\gamma)}(Q_A/Q_B)^{1-\gamma}
{W^{L,T}_A(2-\gamma)W^{L,T}_B(\gamma)\over(2-\gamma)^2\gamma^2}\,\,\,,
\end{equation}
where
\begin{equation}
\label{b9a}
W^T(\gamma)={4-\gamma\over2-\gamma}{\Gamma^4(1+\half\gamma) \over 
\Gamma(2+\gamma)}
{\Gamma(3-\half\gamma)\Gamma(1-\half\gamma)\over\Gamma(4-\gamma)}\,\,,
\end{equation}
\begin{equation}
\label{b9b}
W^L(\gamma)=2{\Gamma^4(1+\half\gamma)\over\Gamma(2+\gamma)}
{\Gamma^2(2-\half\gamma)\over\Gamma(4-\gamma)}\,\, .
\end{equation}
Setting $\xi=s/(cQ_AQ_B)$ we obtain $\sigma_{\gamma\gamma}$ of Ref.
\cite{BHS}. The transition between Eq. (8) of Ref. \cite{BHS} and 
(\ref{b8}) goes through the substitution $\gamma =2\gamma^{\prime}$
which changes $\chi$ of (\ref{chi}) and transforms the r.h.s. of
(\ref{b8}) into the r.h.s. of Eq. (8) in \cite{BHS}, multiplied by a
factor $8\over9$. This factor can be traced to an approximation made
in Ref. \cite{Mc}.

For the sake of completeness we give also the saddle point
approximation of (\ref{b8})

\begin{equation}
\label{b10}
\sigma_{\gamma\gamma} = 
{16\alpha^2 N^2\alpha^2_{em}(e^2_f)^2\over\pi Q_AQ_B}
\xi^{\Delta_p}\sqrt{2a_{\xi}\over\pi}e^{-\half a_{\xi}\log^2(Q_A/Q_B)}
W_A(1)W_B(1)\,\, ,
\end{equation}
where
\begin{equation}
\label{b10a}
W^T(1)={9\pi^3\over256}\,\,\,\,,
W^L(1)={2\pi^3\over256}\,\,\,.
\end{equation}
Note that for the case discussed in Ref. \cite{NPR} we have 
$\xi=s/(cQ^2_>)$ (compare Eq. (\ref{s0b}), and the same set of formulae 
for $\sigma_{\gamma\gamma}$ (\ref{b8}) - (\ref{b10})).

\end{document}